\documentclass[preprint,12pt]{elsarticle}

\usepackage{graphicx}
\usepackage{grffile}

\usepackage{amssymb}

\usepackage{changes}
\usepackage{ulem}
\usepackage{amsmath}
\usepackage{booktabs}
\usepackage{xcolor}
\usepackage{subcaption}
\usepackage{tikz}

\usepackage[ruled,linesnumbered]{algorithm2e}
\newcommand{\RomanNumeralCaps}[1]{\MakeUppercase{\romannumeral #1}}

\begin{document}

\begin{frontmatter}



\title{A machine-learning based method to generate random packed isotropic porous media with desired porosity and permeability}


\author[a,b,c]{Li Jianhui}
\author[a,b,c]{Tang Tingting}
\author[a,b,c,e]{Yu Shimin}
\author[a,b,c]{Yu Peng\corref{d}}

\address[a]{Department of Mechanics and Aerospace Engineering, Southern University of Science and Technology, Shenzhen, 518055, China}
\address[b]{Guangdong Provincial Key Laboratory of Turbulence Research and Applications, Southern University of Science and Technology, Shenzhen, 518055, China}
\address[c]{Center for Complex Flows and Soft Matter Research, Southern University of Science and Technology, Shenzhen 518055, China}
\address[e]{Harbin Institute of Technology, Harbin 150001, China}

\cortext[d]{Corresponding Author}

\begin{abstract}
	Porous materials are used in many fields, including energy industry, agriculture, medical industry, etc. The generation of digital porous media facilitates the fabrication of real porous media and the analysis of their properties. The past random digital porous media generation methods are unable to generate a porous medium with a specific permeability. A new method is proposed in the present study, which can generate the random packed isotropic porous media with specific porosity and permeability. Firstly, the process of generating the random packed isotropic porous media is detailed. Secondly, the permeability of the generated porous media is calculated with the multi-relaxation time (MRT) lattice Boltzmann method (LBM), which is prepared for the training of convolutional neural network (CNN). Thirdly, 3000 samples on the microstructure of porous media and their permeabilities are used to train the CNN model. The trained model is very effective in predicting the permeability of a porous medium. Finally, our method is elaborated and the choice of target permeability in this method is discussed. With the support of a powerful computer, a porous medium that satisfies the error condition of porosity and permeability can be generated in a short time.
\end{abstract}


\end{frontmatter}


\section{Introduction}
\label{Introduction}
A porous medium is a material containing pores or voids \cite{su2012hierarchically}. Permeability is one of the most important parameters for characterizing the transport properties of porous media. The fluid flow through porous material is of great significance in the fields of the energy industry, agriculture, medical industry, etc. For example, in fuel cell, the optimized permeability and wettability of the porous gas diffusion layer, which regulate the transmission of gas and water, could maximize the working efficiency of fuel cell \cite{niu2018two}. In oil and gas exploitation industry, the permeability of underground rocks seriously influences the recovery rate \cite{christensen2001review}. In terms of soil evaporation, the proper temperature and permeability can ensure that the soil maintains a certain humidity \cite{or2013advances}. In the process of the settling of permeable microbial granules in biological wastewater treatment system, the drag coefficient of microbial granules depends heavily on their permeability and porosity \cite{Mu2008Drag}. In food industry, the permeability of porous food affects the speed of food cooling \cite{mellor1980vacuum}. 

Permeability depends on the microstructure of the solid phase in porous media and is independent of the properties of working fluid \cite{nakayama1995pc}. Many studies have attempted to relate permeability with the microstructure and material properties of porous medium \cite{dmitriev1995surface, plougonven2009link, yazdchi_microstructural_2011,ibrahim_microstructure_2019}. The main parameters which affect the permeability of a porous medium are porosity, pore shape (or particle shape), and the interconnection of pores \cite{yazdchi_microstructural_2011,ibrahim_microstructure_2019}. The permeability is very important for analysing porous flow in the representative elementary volume (REV) scale, which should be obtained before the analysis for practical applications. Many models have been proposed for the determination of permeability \cite{xu2008developing}. The Kozeny-Carman equation \cite{kozeny1927uber, carman1937fluid} is the most common model to calculate the permeability, which reads
\begin{equation}
\label{equ:CK}
	k = \frac{\varphi d_h^2}{16K},
\end{equation}
where $\varphi$ is porosity of the porous medium, $K$ is the Kozeny constant, and $d_h$ is the pore hydraulic diameter, which is defined as,
\begin{equation}
	d_h = \frac{4\varphi}{A_0(1-\varphi)}
\end{equation}
where $A_0$ is the ratio of the fluid-solid interfacial area to the solid volume. \par

The Kozeny constant $K$ includes the effects of the flow path (i.e., tortuosity), particle shape, and their connections. Although many studies have been performed for the determination of Kozeny constant \cite{koponen1997permeability, heijs1995numerical,ozgumus_determination_2014} and the calculation of permeability, they are all subject to specific conditions. The result from Koponen \cite{koponen1997permeability} is suitable for the 2-D system consisting of randomly placed identical rectangle obstacles with unrestricted overlap. Anton and Christopher \cite{heijs1995numerical} obtained a good estimate for the permeability of the random array of spheres through the semiempirical Carman-Kozeny equation. Ozgumus \cite{ozgumus_determination_2014} investigated the applicability of Kozeny-Carman equation to the periodic porous medium made from rectangular rods. The Kozeny constant for the porous medium composed of fibers was studied by Kyan \cite{kyan1970flow}, Bechtold and Ye \cite{bechtold2003influence}, Rodriguez \cite{rodriguezRTM2004}, Li and Gu \cite{LI20051}, Chen and Papathanasiou \cite{chen2006variability}, Pacella \cite{pacella2011darcy}, Liu and Hwang \cite{liu2012permeability}. The modified Kozeny-Carman equations for the sandstone were proposed by Mavko \cite{mavko1997effect} and Pape \cite{pape2000}. An analytical method based on the pore scale was also introduced by Arns and Adler \cite{arns_fast_2018} to calculate the permeability of porous medium. The above lieterature suvey indicates that the shape and the arrangement of the basic elmenent for the porous medium have a great influence on the permeability. Therefore, for different types of porous media, the original Kozeny-Carman equation needs to be modified for the permeablity calculation. However, the accuracy for the permeability calculated by the Kozeny-Carman equation depends on empirical parameters and may not meet requirement for high-fidelity analysis.\par

To obtain accurate permeability, experimental measurement in porous media \cite{sharma2008permeability} and numerical simulation in their digitized pairs \cite{song2017new, dongxing2021determination} are two reliable methods. However, the experimental method requires the experimental sample with suitable size and shape, which may not be practical in some situations. On the other hand, the acquisition of digitized porous media requires computerized tomography (CT) or nuclear magnetic resonance (NMR) scanning, which incurs high costs. \par
In addition to the above mentioned real porous medium, digital porous medium, which is a numerical construction of porous medium by different shape of constituent elements, are also widely adopted in the academic studies. On the one hand, the conservation laws that govern fluid flow are the same in both the real and digital porous media \cite{ADLER1990691, CHERBLANC20071127}. The physical properties of the digital porous media, such as porosity, tortuosity, skeletal morphology, etc., can be very similiar to those of the real porous media  \cite{mosser_reconstruction_2017, shams_hybrid_2021, nguyen_synthesizing_2022, zhang_3d_2022}. The results based on digital porous media have been successfully applied in industries such as oil and gas \cite{FOROOZESH2020113876, wu20063d}. On the other hand, in the field of digital twins and material fabrication, the design of digital porous media in advance can guide the fabrication of porous materials in practical applications \cite{ma12030541}. \par

The digital porous media have been designed and applied in different investigations by many researchers. Circles \cite{tang2019investigation,Tang2020}, shuttles \cite{barta2006creeping}, and rectangles \cite{koponen_permeability_1997} of the same size are regularly arranged to represent porous media. Circles with different radius are also widely applied to construct porous media, e.g., Zakirov \cite{zakirov_prediction_2020}, Borgman \cite{borgman_immiscible_2019}, and Holtzman \cite{holtzman_effects_2016} used the disorder parameter to constrain the circle packed porous media. Besides, completely random methods have been proposed to generate porous media. The quartet structure generation set (QSGS) method can generate random porous media by defining porosity, growth core distribution probability, and directional growth probability \cite{Wang2007}. By adjusting related parameters, this method can generate two-dimensional or three-dimensional and isotropic or anisotropic porous media. The simulated annealing (SA) method also can generate random porous media by defining porosity and two-point probability function of solid phase in porous media \cite{Yeong1998495, Liu2017}. Gaussian random field has been used by Adler \cite{adler_flow_1990} to construct random porous media, with a given porosity and a correlation function. Although the permeability of porous media with regular geometric arrangement can be accurately calculated using the fitting formula \cite{tang2019investigation,zakirov_prediction_2020}, its practical application scenarios are very limited. To obtain the accurate permeability of randomly generated digital porous media, the calculation of its internal flow field has to be involved, which requires a lot of computing costs, especially for three-dimensional situation.\par

In recent years, the machine learning methods, e.g., convolutional neural network (CNN), deep neural network (DNN), etc., have been introduced to predict the transport properties in porous media \cite{Alqahtani2020, Wu2019, Tian2020}. Alqahtani et al. \cite{Alqahtani2018, Alqahtani2020} utilized CNN to rapidly predict the properties from 2D greyscale micro-computed tomography images of the porous media in a supervised learning frame. By using a dataset of greyscale micro-CT images of three different sandstone species as the input data, the porosity, coordination number, and average pore size are predicted. Wu et al. \cite{Wu2018} concluded that the fast prediction of permeability directly from images enabled by image recognition neural networks was a novel pore-scale modeling method. They conducted a comparison of machine learning results and the ground truths, which suggested excellent predictive performance across a wide range of porosity and pore geometries, especially for those with dilated pores. Meanwhile, computational time was reduced by several orders of magnitude compared to that for compuational fluid dynamics (CFD) simulations. Wu et al. \cite{Wu2019} reported the application of machine learning methods for predicting the effective diffusivity of two-dimensional porous media from their structure images. The effective diffusivity of porous media was computed by lattice Boltzmann method (LBM), which were used to train CNN models and evaluate their performance. The trained model predicts the effective diffusivity of porous structures with computational cost orders of magnitude lower than that for LBM simulations and provides better prediction than that using the empirical Bruggeman equation. Tian et al. \cite{Tian2020} proposed an artificial neural network (ANN) and genetic algorithm (GA) hybrid machine learning model to implicitly build a nonlinear relationship between pore structure parameters and permeability. The prediction results showed that the ANN-GA model was robust in predicting permeability based on pore structure parameters.  \par
Generally, it is easy to constrain the porosity by setting the control parameter when generating the digital porous medium. However, the accurate permeability of the digital porous dedia has to be calculated by using CFD simulation, which may be time-consuming in three-dimensional cases. If sufficent data can be obtained from simulations, the machine learning model may be trained and applied to rapidly perdict the permeablity. This means that the permeability can only be dertermined after of the digital porous medium has been generated. However, in pratical applications, the porous media with desired properties (e.g., porosity and permeability) are often required. This motivates that the present study to develop a random generation method which can construct digital porous medium with given porosity and permeablity.  \par

In this paper, an algorithm to generate the random packed isotropic porous medium is proposed. The MRT-LBM based on the D2Q9 scheme is applied to simulate the internal flow of the digital porous medium, based on which the permeability can be evaluated. Massive simulations are performed and the collective data are then used to train the CNN model. Based on the porous media generation method and the trained CNN model, a method for the generation of random packed isotropic porous media with specific porosity and permeability is proposed. \par

The remaining of this paper is organized as follows. The generation method  for random packed isotropic porous media is proposed in Section \ref{GoPM}. The details on the numerical simulations based on MRT-LBM-D2Q9 scheme for the internal flow in porous media are described in Section \ref{CP}. The training of the CNN model is presented in Section \ref{CNN}. Then the method for the generation of random packed isotropic porous media with specific porosity and permeability is detailed in Section \ref{method}. Finally, the conclusions are drawn in Section \ref{conclusion}.

\section{Generation of porous media}
\label{GoPM}
The porous medium considered in this study is packed with circles of different radius, which follow the lognormal distribution \cite{perkins1963review}. The probability density function (PDF) of lognormal distribution is shown as
\begin{equation}
\label{lognormal}
	PDF = \frac{1}{ r\sigma\sqrt{2 \pi}} e^{-(\ln r - \mu)^2/2\sigma^2}, r \geq 0,
\end{equation}
where $r$ is the radius, $\mu$ is the expected value of all radii's natural logarithm, $\sigma$ is the standard deviation of all radii's natural logarithm. After the parameters of the PDF equation are determined, the number of circles $m$ and the porosity $\varphi$ of the porous media are negatively correlated in a region of a specific size, which can be expressed as:
 \begin{equation}
	\label{equ:rhoml}
	\varphi = 1 - \frac{S(m)}{L^2},
 \end{equation}
 where $L$ is the side length of the square area, $S$ is a function of $m$ and represents the sum of the areas of all circles. During the generation of porous media, the discrete distribution of circle radii will be determined by the following algorithm. Firstly the target porosity, the absolute error of porosity, and the maximum value ($m_{max}$) and minimum value ($m_{min}$) of the number of circles need to be determined in advance. Then a target value $m$ is guessed between $m_{max}$ and $m_{min}$. $m$ discrete circle radii are generated by the PDF. Porosity is calculated from the generated circle radii. This porosity is compared with the target porosity and the size of $m$ is adjusted according to the comparison result. Repeat the previous steps until the calculated porosity meets the error requirement with the target porosity. At this point a discrete distribution of suitable circle radii is obtained. The specific algorithm is shown in Algorithm \ref{algorithm}, where we use a multi-threaded environment to speed up the implementation of the algorithm. Before implementing Algorithm \ref{algorithm}, the rough values of $m_{max}$ and $m_{min}$ can be obtained using program debugging based on Equations \ref{lognormal} and \ref{equ:rhoml}, and the determined $L$ and $\varphi$. Or first determine the values of $m_{max}$ or $m_{min}$ and $\varphi$, and then get the value of $L$ by program debugging. \par

\begin{figure}[!htbp]
    \centering
	\includegraphics[width=0.8\linewidth]{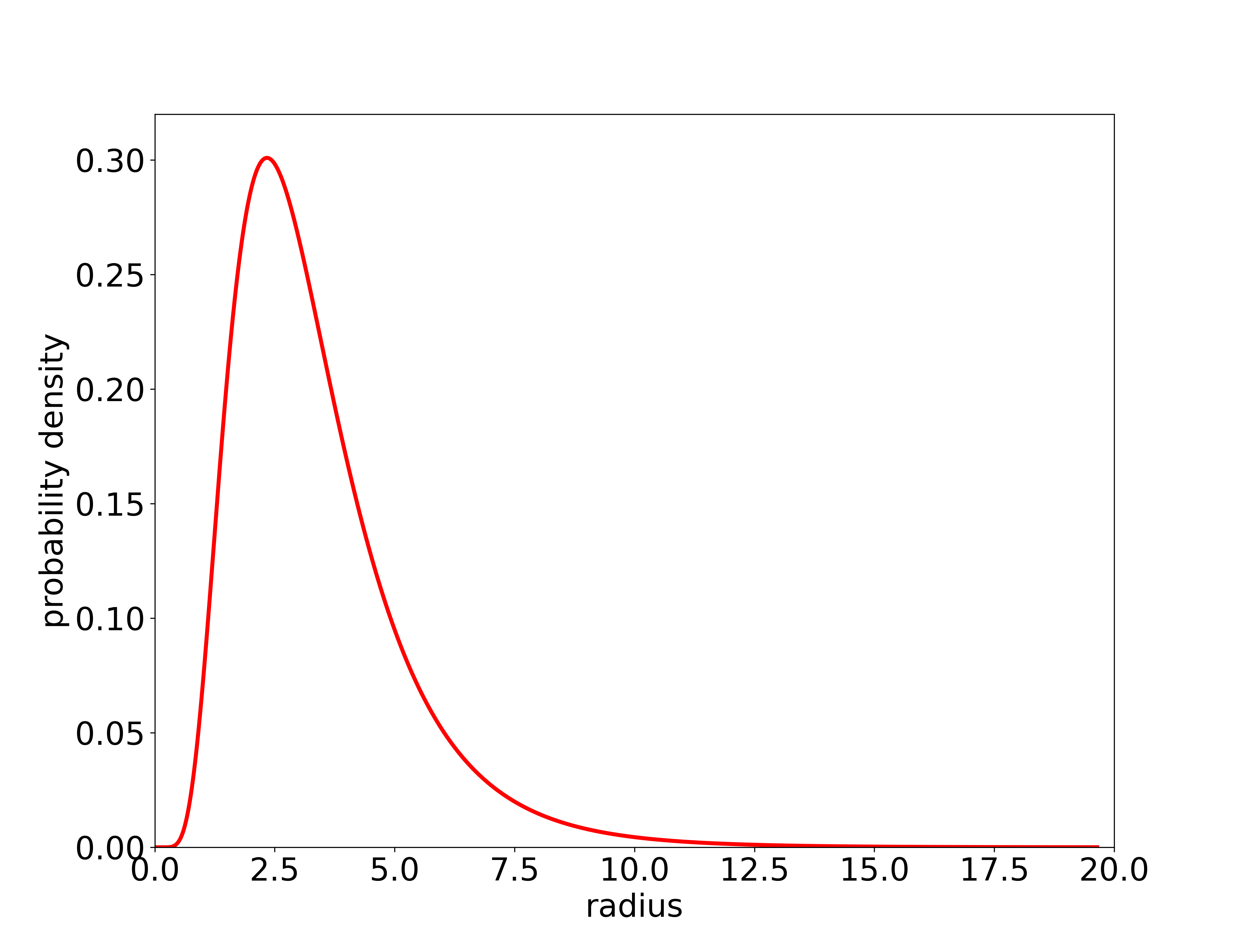}
	\caption{The probability density distribution of circle radius.}
	\label{fig:pd}
\end{figure}

In this study, $\mu$ and $\sigma$ are specially set to be 1.1 and 0.5, respectively. The size of porous area is $L^2$, $L = 200$. The probability density distribution of circle radius is shown in Figure \ref{fig:pd}. The porosity corresponding to $m_{max} = 300$ is about 0.3, and to $m_{min} = 25$ is about 0.94. The discrete circle radii are generated using the lognormal distribution function of the C++ standard library. \par

\begin{algorithm}
	\caption{Determination of circles number and radius}
	\label{algorithm}
	\KwIn{target porosity $\varphi$, the specified tolerance for porosity $\xi$ }
	\KwData{$m_{max}$, $m_{min}$, guessed circles number $m$ ($m_{min} < m < m_{max}$)}
	\KwResult{A vector of circles radius $V$, the porosity $\varphi'$ of the generated porous medium}
	Get the thread number of processor $N$\;
	\While{True}{
		In thread $i$, a vector $V_i$ is generated randomly, whose values follow the lognormal distribution. The vector size is $m$\;
		Calculate the intermediate porosity $\varphi_i$ correspongding to vector $V_i$\;
		Calculate the number of threads $N'$, where $\varphi_i > \varphi$\;
		\eIf{$N' \ne (N-N')$ when N is even, $N' \ne (N-N' \pm 1)$ when N is odd}{
			\eIf{$N' > (N-N')$ when N is even, $N' > (N-N'+2)$ when N is odd}{
				$m_{min} = m$\;
				$m = (m_{max} + m_{min})/2$\;
			}{
				$m_{max} = m$\;
				$m = (m_{max} + m_{min})/2$\;
			}
		}{
			\For{each thread $i$}{
				\If{the absolut value of $\varphi - \varphi_i$ is less than $\xi$}{
					$\varphi' = \varphi_i$\;
					$V = V_i$\;
					Exit loop\;
				}
			}
		}
	}	
\end{algorithm}
\par
\begin{figure}
	\centering
    \begin{subfigure}{\textwidth}
		\centering
		\includegraphics[width=0.4\linewidth]{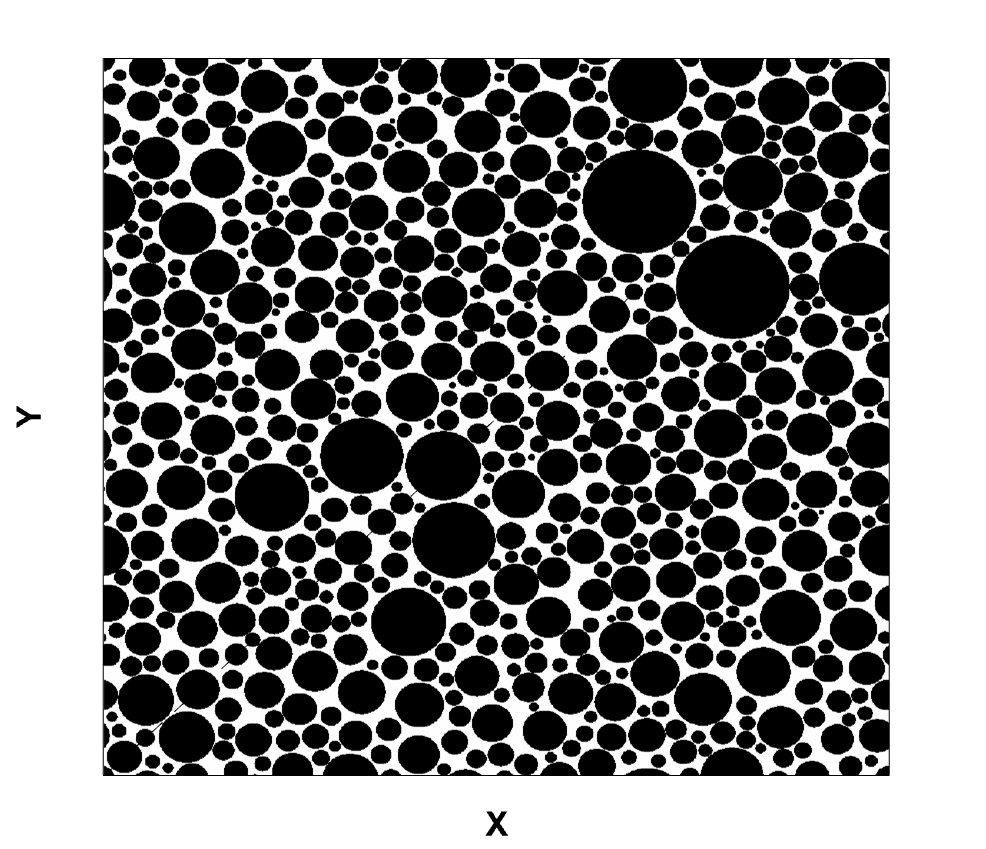}
		\caption{$\varphi = 0.3$}
		\label{fig:phi03}
	\end{subfigure}
	\begin{subfigure}{\textwidth}
		\centering
		\includegraphics[width=0.4\linewidth]{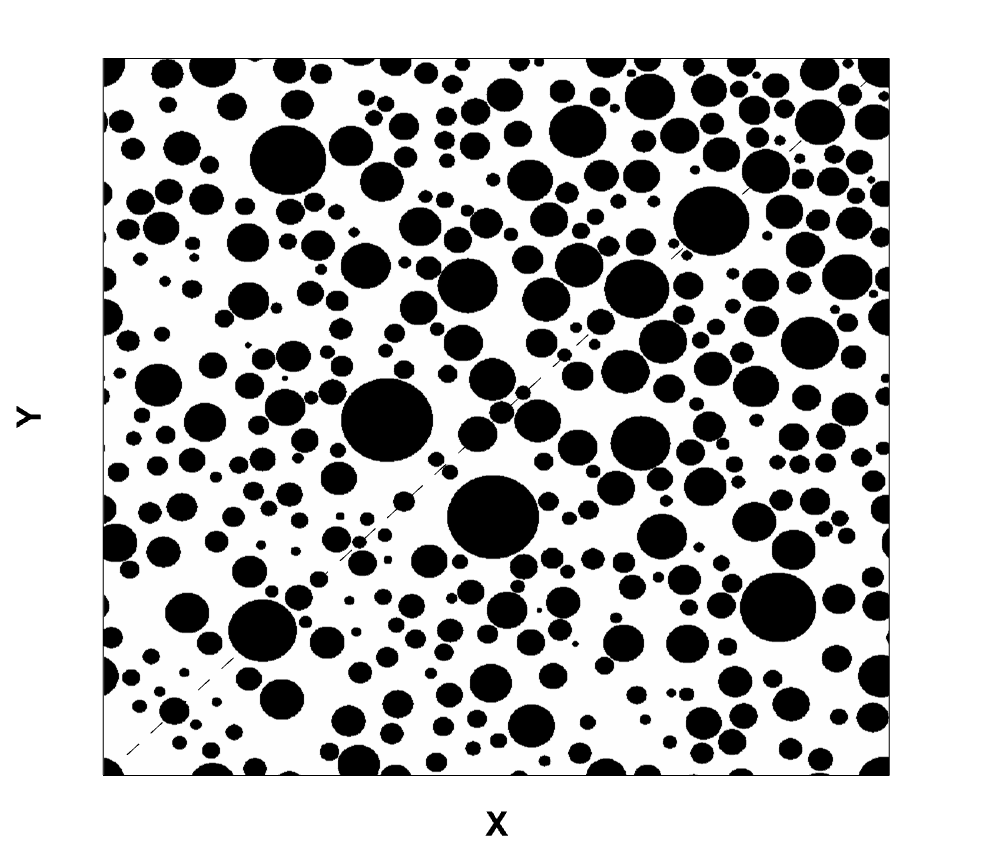}
		\caption{$\varphi = 0.6$}
		\label{fig:phi06}
	\end{subfigure}
	\begin{subfigure}{\textwidth}
		\centering
		\includegraphics[width=0.4\linewidth]{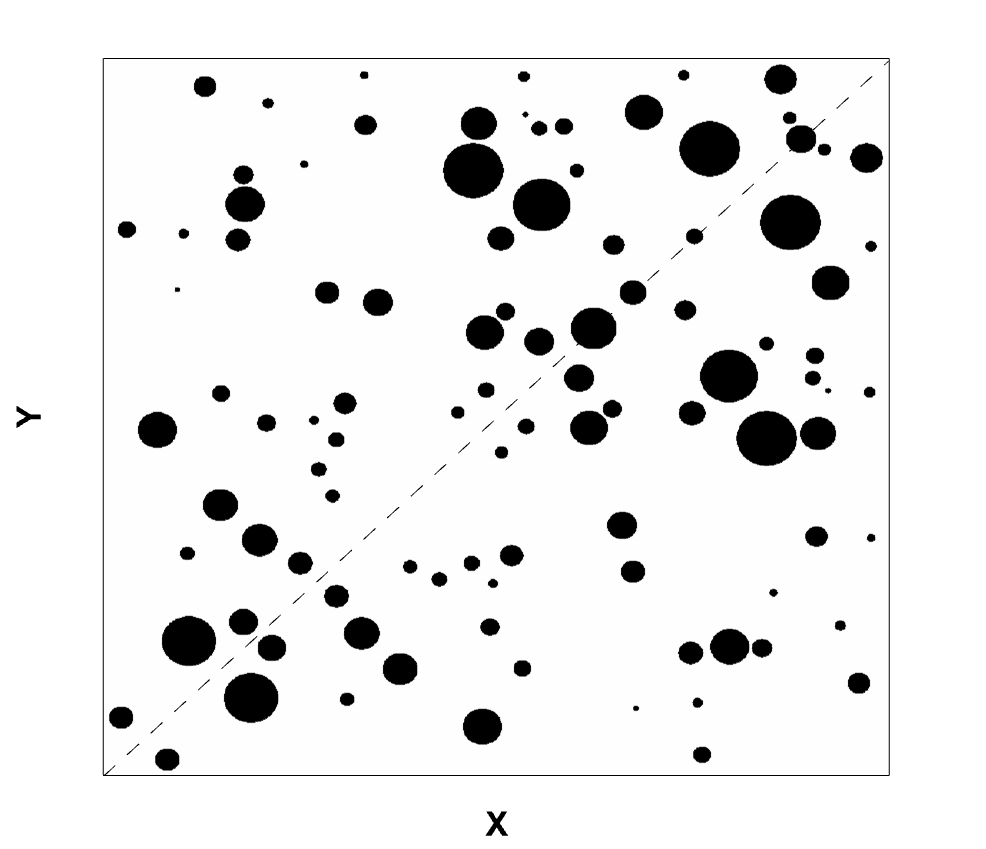}
		\caption{$\varphi = 0.9$}
		\label{fig:phi09}
	\end{subfigure}
	\caption{Porous media with different porosities.}
	\label{fig:phi}
\end{figure}

After circles number and radius are determined, all circles will be filled into the porous area in order of radius from the largest to smallest. The position of each circle is random. The minimum distance between every two circles is 0.1. To ensure the isotropic properties of porous media, all circles will be filled symmetrically with $y = x$ (the dashed line in Figure \ref{fig:phi}) as the axis of symmetry. This method keeps the flow field formed by the flow coming from both X and Y directions in the porous medium the same. A multi-threaded calculation can speed up the filling speed. In this study, it takes less than 200 milliseconds to generate a porous media with $\epsilon = 0.001$ and $\varphi = 0.55$ under the condition of eight threads. Figure \ref{fig:phi} shows three porous media with $\varphi = $ 0.3, 0.6 and 0.9, respectively.\par

At the end of the generation, the entire porous area will be divided into $1024 \times 1024$ lattices for LBM calculations and CNN training. During this conversion process, the jagged boundary replaces the curved boundary, which is usually used in the studies of flow in porous media \cite{zakirov_prediction_2020,askari_thermal_2015,santos_computationally_2021}. 

\section{Calculation of permeability}
\label{CP}
The permeability of the porous media generated in section \ref{GoPM} is calculated using the Lattice Boltzmann Method (LBM). The value of permeability will be used as the labeled data for training the CNN model in section \ref{CNN}. \par

\subsection{The MRT-LBM-D2Q9 model}
In this study, the flow in porous media is simulated by LBM using the multi-relaxation time (MRT) scheme \cite{noauthor_generalized_1992, lallemand_theory_2000, wu_simulation_2004,yang_mrt_2012, aslan2014investigation}. The LBM-MRT collision model of nine velocities on a two-dimensional lattice is given by:
\begin{equation}
	\mathbf{f}\left( \vec{x} + \vec{e} \Delta t, t + \Delta t \right) - \mathbf{f} \left( \vec{x}, t \right) = -M^{-1} S \left[ \mathbf{m} \left( \vec{x}, t \right)  - \mathbf{m}^{eq}\left( \vec{x}, t \right) \right],
\end{equation}
where $t$, $\Delta t$ and $\vec{x}$ represent the time, time step, and particle spatial position, respectively. \par

The discrete velocity vector $\vec{e}$ for D2Q9 is given by:
\begin{equation}
    \vec{e} = c \begin{bmatrix}
    0 & 1 & -1 & 0 & 0 & 1 & -1 & -1 &  1 \\
    0 & 0 & 0 & 1 & -1 & 1 & 1 & -1 & -1
    \end{bmatrix},
\end{equation}
where $c = \Delta x/\Delta t$ is the lattice speed, $\Delta x$ is the lattice length and $\mathbf{f}$ is the column vector of the particle distribution, expressed as:
\begin{equation}
    \mathbf{f} = (f_0, f_1, f_2, f_3, f_4, f_5, f_6, f_7, f_8)^T,
\end{equation}
where $T$ is the transpose operator. \par

$M$ is a $9 \times 9$ matrix which linearly transforms the distributions functions $\mathbf{f}$ to the moment $\mathbf{m}$:
\begin{equation}
    \mathbf{m} = M \mathbf{f}, \mathbf{f} = M^{-1} \mathbf{m}
\end{equation}
\begin{equation}
M = \begin{bmatrix}
1& 1 & 1 & 1 & 1 & 1 & 1 &  1& 1\\ 
-4& -1 & 2 & -1 & 2 & -1 & 2 & -1 & 2\\ 
4& -2 & 1 & -2 & 1 & -2 & 1 & -2 & 1\\ 
0& 1 & 1 & 0 & -1 & -1 & -1 & 0 & 1\\ 
0& -2 & 1 & 0 &  -1&  2&  -1&  0& 1\\
0& 0 & 1 & 1 & 1 & 0 & -1 & -1 & -1\\ 
0& 0 & 1 & -2 & 1 & 0 & -1 & 2 & -1\\ 
0& 1 & 0 & -1 & 0 & 1 & 0 & -1 & 0\\ 
0& 0 & 1 & 0 & -1 & 0 & 1 & 0 & -1
\end{bmatrix}
\end{equation}

For the D2Q9 lattice model, $\mathbf{m}$ is a column vector of macroscopic variables:
\begin{equation}
\mathbf{m} = (\rho, e, \epsilon, j_x, q_x, j_y, q_y, p_{xx}, p_{xy})^T,
\end{equation}
where $\rho$ is the fluid density, $\epsilon$ is related to the square of the energy $E$, $j_x = \rho u_x$ and $j_y = \rho u_y$ are respectively the mass flux in two directions, $q_x$ and $q_y$ correspond to the energy flux in two directions, and $p_{xx}$ and $p_{xy}$ correspond to the diagonal and off-diagonal component of the viscous stress tensor. $\rho$, $j_x$, and $j_y$ are the conserved moments in the system. Other moments are non-conserved moments and their equilibria are functions of the conserved moments in the system \cite{noauthor_generalized_1992, lallemand_theory_2000}. All elements of $\mathbf{m}^{eq}$ can be calculated as:
\begin{equation}
    \begin{array}{rl}
    m_0^{eq} = & \rho \\
    m_1^{eq} = &-2 \rho + 3(j_x^2 + j_y^2) \\
    m_2^{eq} = &\rho - 3(j_x^2 + j_y^2) \\
    m_3^{eq} = &j_x \\
    m_4^{eq} = &-j_x \\
    m_5^{eq} = &j_y \\
    m_6^{eq} = &-j_y \\
    m_7^{eq} = &j_x^2 - j_y^2  \\
    m_8^{eq} = &j_x j_y 
    \end{array}.
\end{equation}
$S$ is a non-negative $9 \times 9$ diagonal relaxation matrix expressed as:
\begin{equation}
    S = diag (s_0, s_1, s_2, s_3, s_4, s_5, s_6, s_7, s_8).
	\label{eq:S}
\end{equation}
In Equ. \ref{eq:S}, $s_0 = s_3 = s_5 = 0$ enforces mass and momentum conservation before and after collision. Lallemand and Luo\cite{lallemand_theory_2000} showed that the MRT model can reproduce the same viscosity as that of the single-relaxation time (SRT) model if $s_7 = s_8 = 1/\tau$, where $\tau$ is the collision relaxation time in SRT model. The rest of the relaxation parameters ($s_1, s_2, s_4, s_6$) can be more flexibly chosen. \par
At the beginning of the calculation, the particle distribution is initialized with the equilibrium density distribution function $f_i^{eq}$ in the SRT model, which is computed as:
\begin{equation}
f_i^{eq} = \omega_i \rho \left[ 1 + \frac{\vec{e}_i \cdot \vec{u}}{c_s^2} + \frac{ \left( \vec{e}_i \cdot \vec{u}\right)^2}{2 c_s^4}  - \frac{u^2}{2 c_s^2}\right],
\end{equation}
where $\omega_i$ is the weight coefficient defined by:
\begin{equation}
\omega_i = \left \{ 
\begin{array}{ll}
4/9	& i = 0 \\ 
1/9 & i = 1,2,3,4 \\ 
1/36& i = 5,6,7,8 
\end{array} \right . .
\end{equation}
\par
The fluid density $\rho$ and velocity $\vec{u}$ are computed from the density distribution function $f_i$:
\begin{equation}
\rho = \sum_{i}f_i^{eq}.
\end{equation}
\begin{equation}
\rho \vec{u} = \sum_{i} \vec{e}_i f_i^{eq}.
\end{equation}
\par
The mass and momentum equations can be derived from the model via multi-scaling expansion \cite{Guo2000} as:
\begin{equation}
\frac{\partial \rho}{\partial t} + \nabla \cdot (\rho \mathbf{u}) = 0,
\end{equation}
\begin{equation}
\frac{(\rho \mathbf{u})}{\partial t} + \nabla \cdot (\rho \mathbf{u} \mathbf{u}) = -\nabla \rho + \nu \left[ \nabla^2(\rho \mathbf{u}) + \nabla(\nabla \cdot (\rho \mathbf{u})) \right],
\end{equation}
where $p$, $c_s$ and $\nu$ are the pressure, sound speed, and kinematic viscosity, respectively:
\begin{equation}
p = c_s^2 \rho,
\end{equation}
\begin{equation}
c_s = c / \sqrt{3},
\end{equation}
\begin{equation}
\nu = (2 \tau - 1)c^2 \Delta t / 6.
\end{equation}
\par
Due to the locality of collision in the LBM, it is highly suitable for parallel computing. In the proposed program, the collision and streaming process runs on the GPU. The initialization of variables and the reading and output of data run on the CPU. \par

\subsection{Verification of LBM program}
The well-known square cavity flow case is applied to verify the accuracy of the proposed LBM program. The lattice size of the square cavity is $400 \times 400$. The velocity in the x-direction of the top cover is 0.1 in lattice units, which is implemented using the Ladd’s method \cite{ladd_1994}. The other three boundaries are no-slip wall condition, which is implemented with a bounce-back scheme. \par

\begin{figure}[!htbp]
    \centering
	\includegraphics[width=0.8\linewidth]{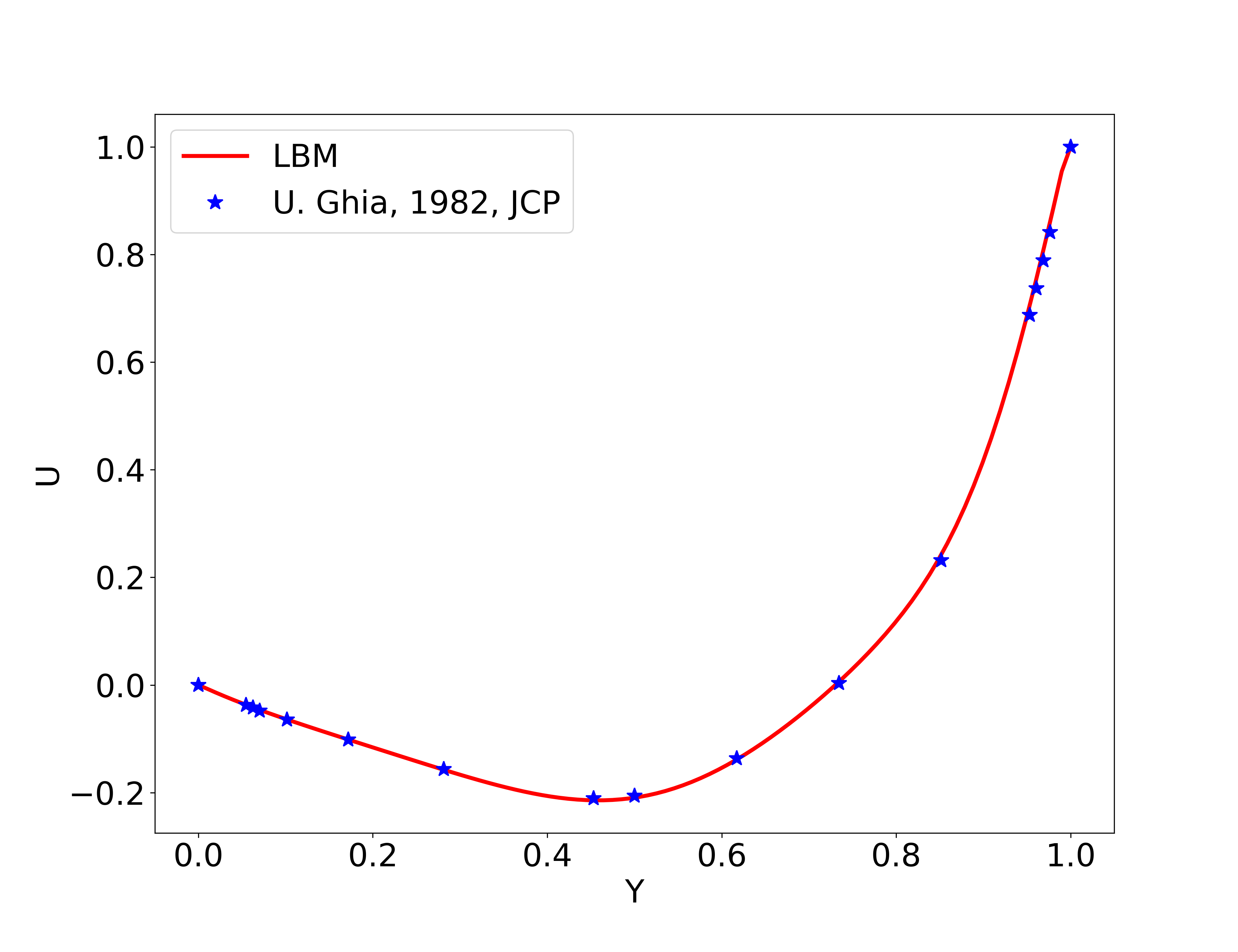}
	\caption{Comparison of the $u_x$ velocity with the results of the literature along the horizontal line through the geometric center of cavity.}
	\label{fig:cwlr}
\end{figure}

In the comparison, the Reynolds number is 100. The results for $u_x$ velocity along the horizontal line through the geometric center of the cavity, are shown in Figure \ref{fig:cwlr}. The x-axis represents the dimensionless position, and the y-axis represents the dimensionless $u_x$ velocity. It can be seen that the obtained results are very consistent with the literature results \cite{ghia1982high}. 
\subsection{Calculation of permeability}
\begin{figure}[!htbp]
    \centering
	\includegraphics[width=0.5\linewidth]{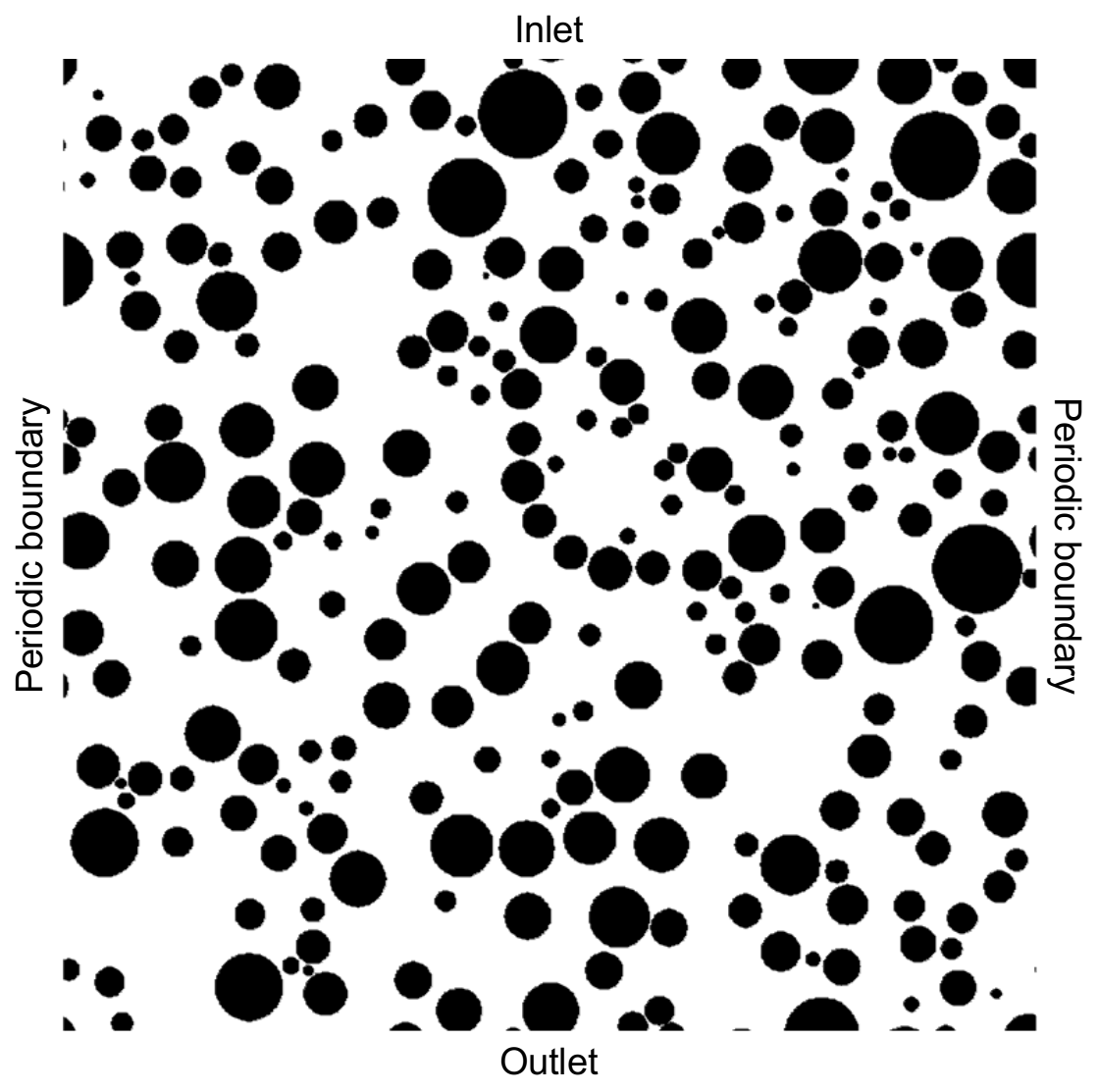}
	\caption{Boundary conditions in LBM calculation.}
	\label{fig:BC}
\end{figure}

Figure \ref{fig:BC} shows the generated porous media and boundary conditions in LBM calculation. The black part in Figure \ref{fig:BC} is the solid area, while  the white part is the fluid area. The inlet and outlet are pressure conditions, that are imposed by specifying the fluid densities at both boundaries. The density difference is set to 0.0001. The non-equilibrium extrapolation method \cite{Guo2002366} is implemented for pressure condition. The left and right boundaries are periodic conditions. The non-slip wall boundary condition is used between fluid and solid.  \par

In this study, the maximum Reynolds number $Re$ is 4 and the characteristic length is $L$. Therefore, the permeability of the random porous media can be calculated by Darcy’s law. The unit of permeability is the international standard (IS) unit ($m^2$). The permeability $K$ can be calculated as:
\begin{equation}
	K = \frac{\overline{U_y}_{LB} \nu_{LB} \rho_{LB}}{dp_{LB}/L_{LB}} \left( \frac{\nu_{real} c_{s,LB}}{\nu_{LB} c_{s,real}} \right)^2,
\end{equation}
where $\overline{U_y}_{LB}$ is the average velocity in $y$ direction in LBM calculation, $L_{LB}$ is the total length of the direction of pressure difference in LBM calculation. \par

In this study, air properties at an atmospheric pressure of $15^{\circ}C$ are used for unit conversion. Note that $\nu_{real} = 1.46 \times 10^{-5} m^2/s$ is the real kinetic viscosity of air and $c_{s,real} = 340.3 m/s$ is the sound speed in air.\par

\section{Training of CNN model}
\label{CNN}
\begin{figure}[!htbp]
    \centering
	\includegraphics[width=\linewidth]{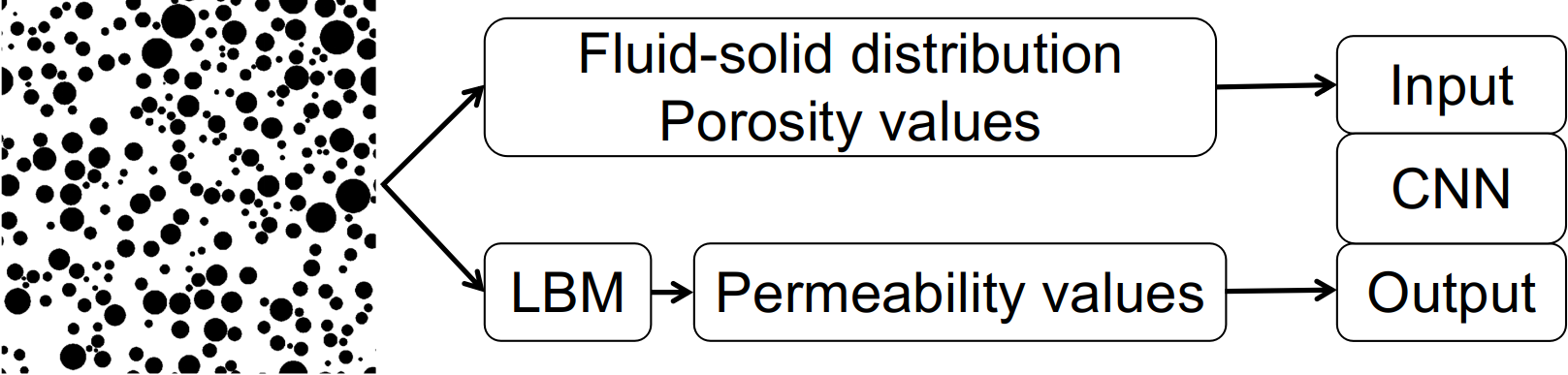}
	\caption{Relationship between section \ref{GoPM}, \ref{CP} and \ref{CNN}.}
	\label{fig:relationship}
\end{figure}

In this section, a CNN model is developed to predict the permeability of the random porous media. The relationship between sections \ref{GoPM}, \ref{CP} and \ref{CNN} is shown in Figure \ref{fig:relationship}. Based on the random porous media generated by the proposed method in section \ref{GoPM}, the images and porosity values of porous media are transported to the input layer of the CNN model. The permeability values calculated by LBM are transported to the output layer of the CNN model. With these data, the CNN model will learn the relationship between the input data and output data.\par

\subsection{The pre-processing of training data} 
\begin{figure}[!htbp]
    \centering
	\includegraphics[width=\linewidth]{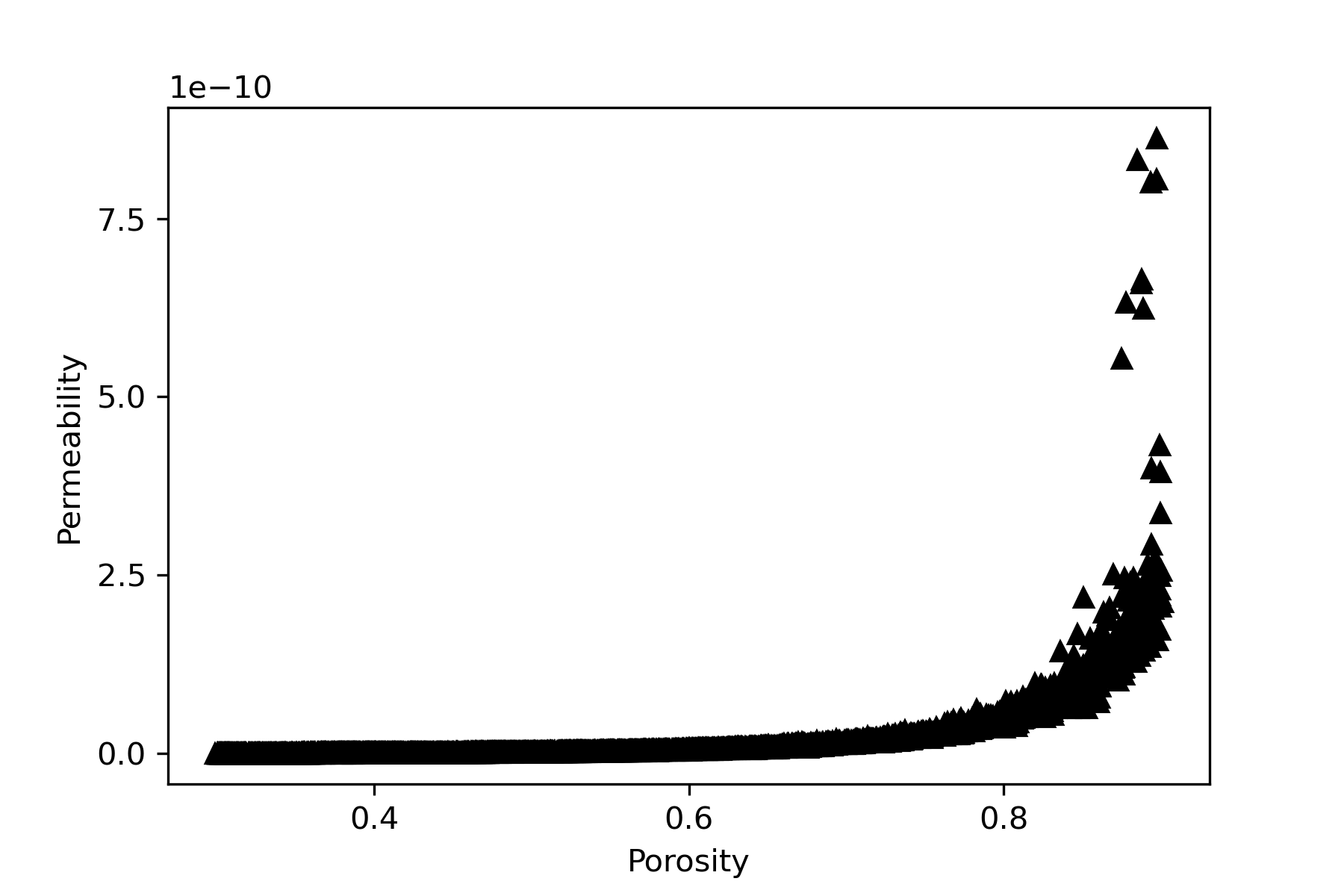}
	\caption{Presentation of all data on the porosity and permeability axes.}
	\label{fig:KPor}
\end{figure}

\begin{figure}[!htbp]
    \centering
    \begin{subfigure}{\textwidth}
	\includegraphics[width=\linewidth]{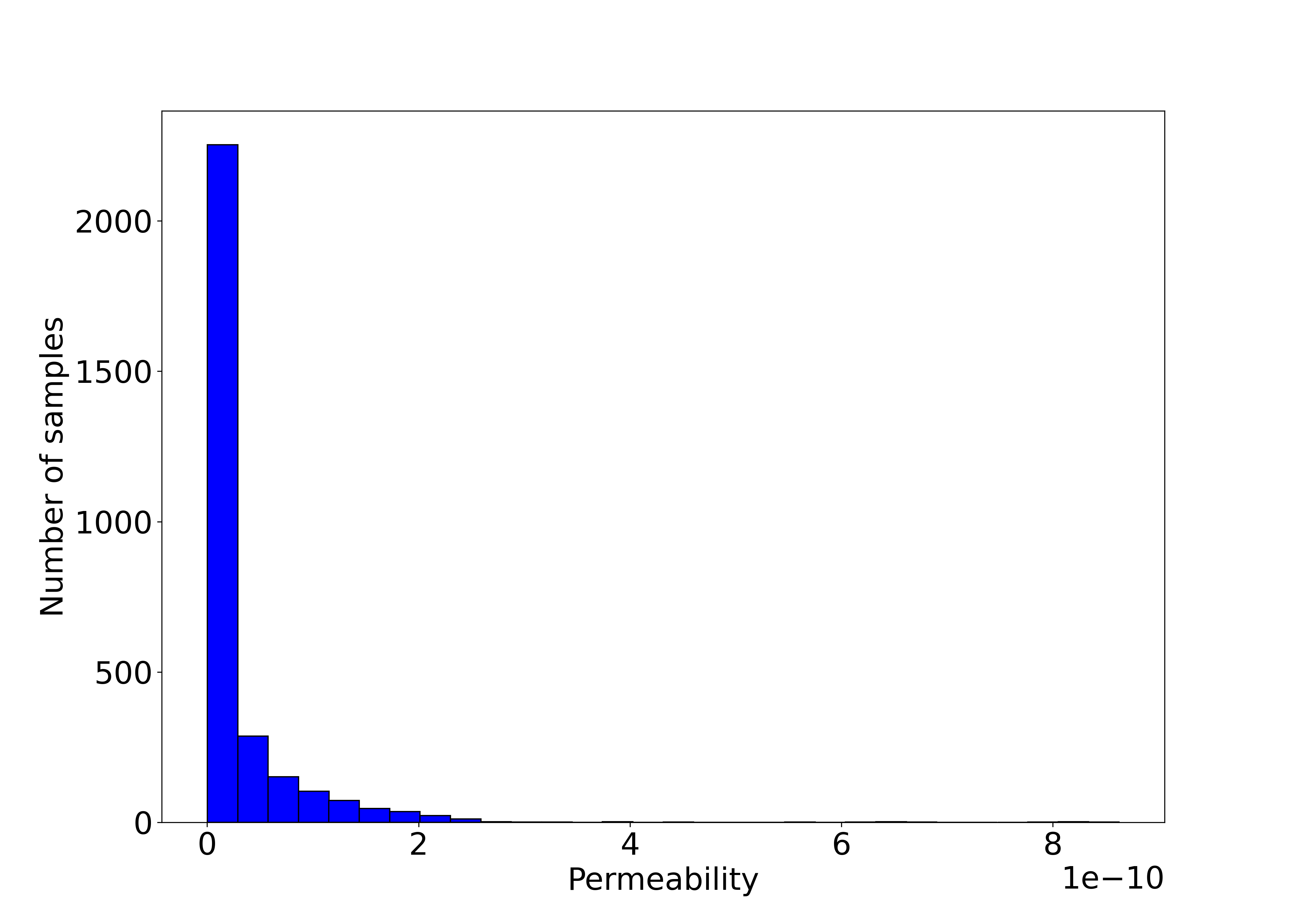}
	\caption{}
	\label{fig:dis_be_A}
	\end{subfigure}
	\begin{subfigure}{\textwidth}
	\includegraphics[width=\linewidth]{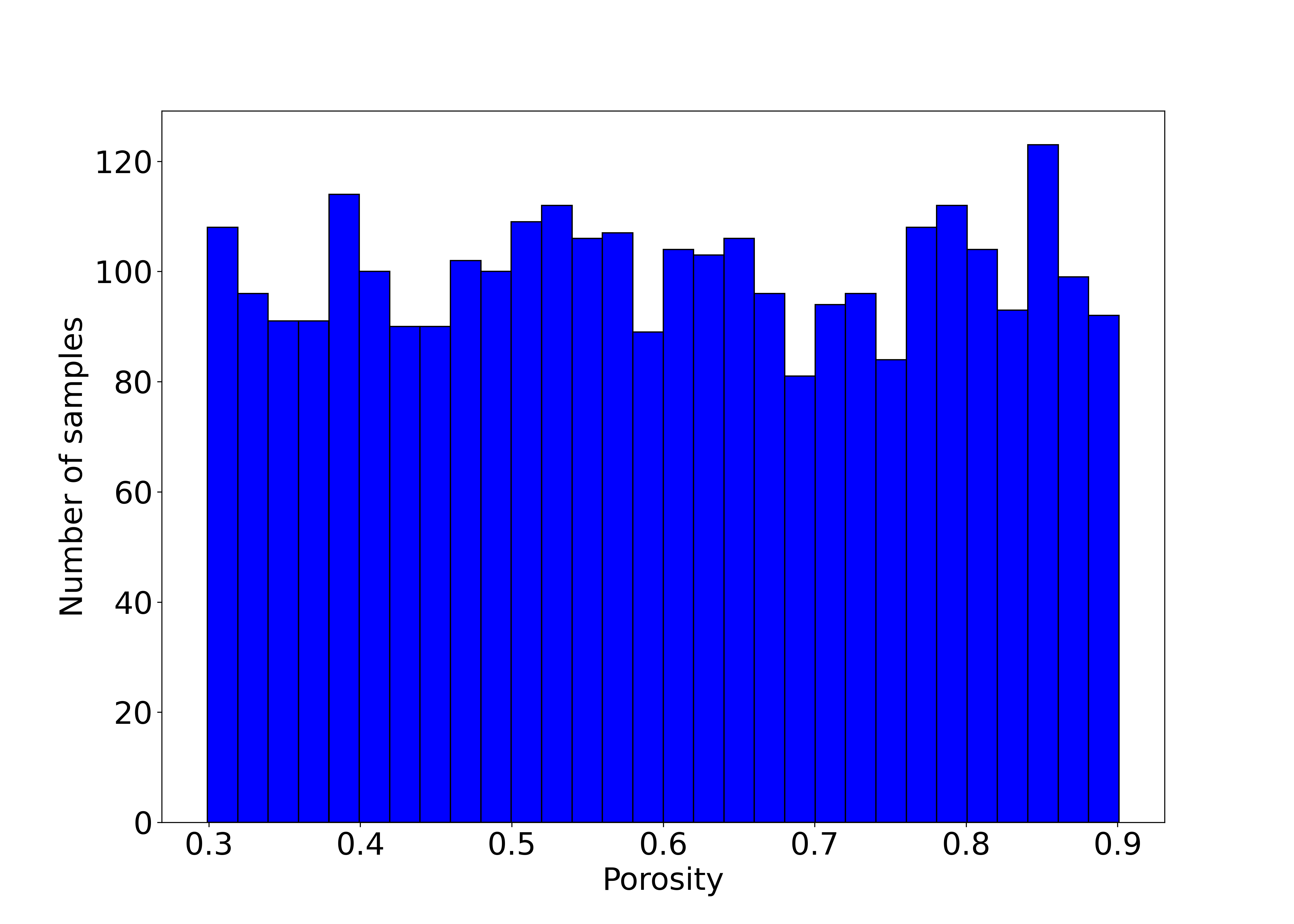}
	\caption{}
	\label{fig:dis_be_B}
	\end{subfigure}
	\caption{The statistical distribution of permeability and porosity values.}
	\label{fig:distribution_beforeNormed}
\end{figure}

In the proposed model, the fluid-solid distribution array size of a porous medium is the same as the LBM Syntheticlattice size, $1024 \times 1024$. Because the representation of porous media is binary, values of the distribution array are zeros and ones. Zero means fluid and one means solid. 3000 samples are generated with a porosity between 0.3 and 0.9. The values of permeability are between $1.57 \times 10^{-13}$ and $8.63 \times 10^{-10}$ $m^2$. The porosity and permeability values of all data are shown in Figure \ref{fig:KPor}. It can be seen that the porosity and permeability are an approximately exponential relationship. The statistical distribution of permeability and porosity values are shown in Figure \ref{fig:distribution_beforeNormed}. It can be concluded from Figure \ref{fig:distribution_beforeNormed} that the number of samples for each porosity is almost the same. However, the number of samples distributed according to permeability is uneven. The number of samples with low permeability is much larger than the number of samples with large permeability. In the process of CNN training, the whole data is split into three datasets, including training, validation and testing. The training dataset is used to fit the model. The validation dataset is used to provide an unbiased evaluation of a model fit on the training dataset while tuning the model’s hyperparameters. The testing dataset is used to provide an unbiased evaluation of a final model fit on the training dataset. In this study, 72\% of the samples are used for training, 18\% for validation and 10\% for testing. \par

\begin{figure}[!htbp]
    \centering
	\includegraphics[width=\linewidth]{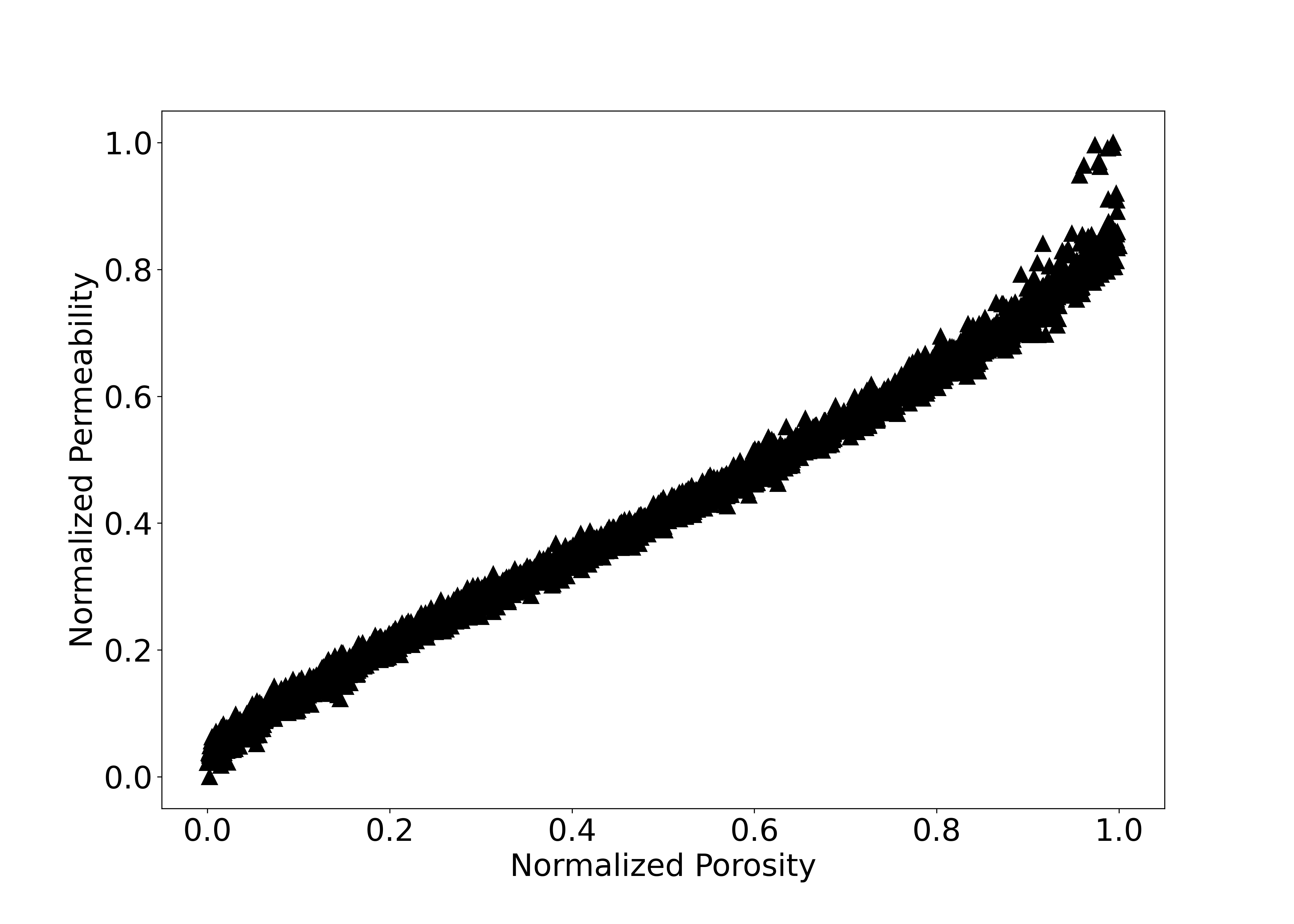}
	\caption{Presentation of all data on the normalized porosity and permeability axes.}
	\label{fig:KPorNormed}
\end{figure}

\begin{figure}[!htbp]
    \centering
	\begin{subfigure}{\textwidth}
	\includegraphics[width=\linewidth]{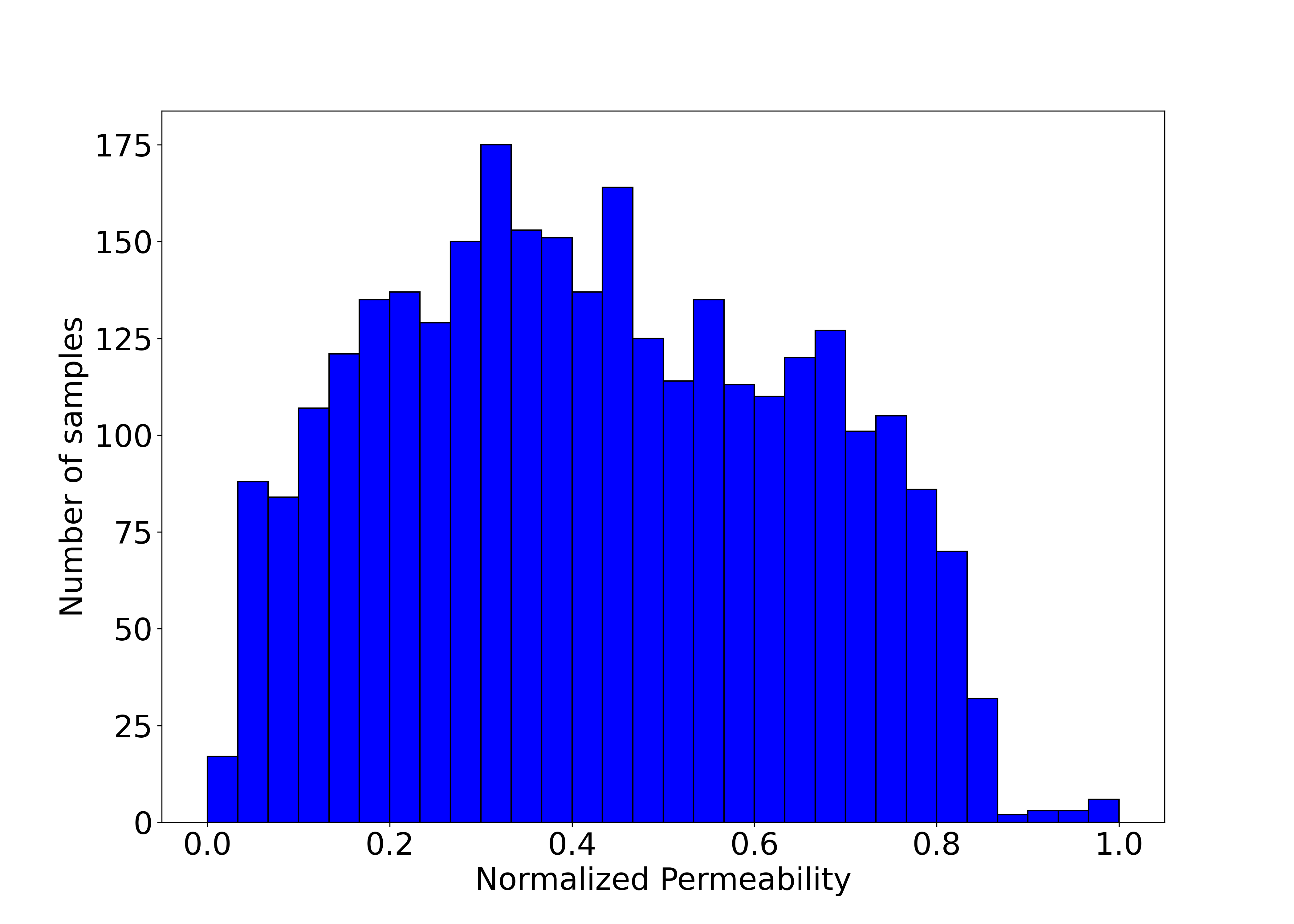}
	\caption{}
	\label{fig:dis_af_A}
	\end{subfigure}
	\begin{subfigure}{\textwidth}
	\includegraphics[width=\linewidth]{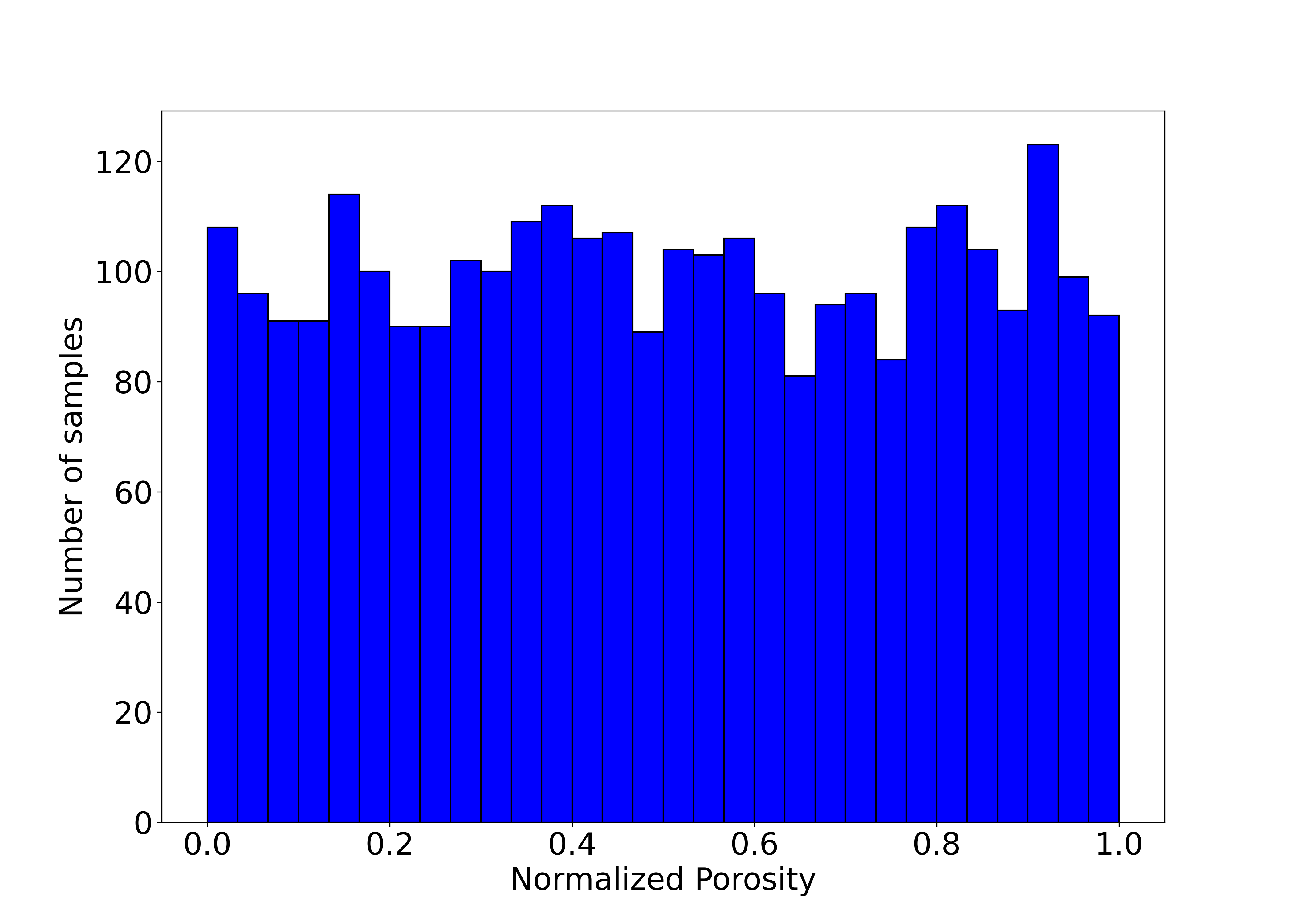}
	\caption{}
	\label{fig:dis_af_B}
	\end{subfigure}
	\caption{The statistical distribution of normalized permeability and porosity values.}
	\label{fig:distribution_afterNormed}
\end{figure}

To stabilize the training process, all permeability and porosity values need to be standardized with min-max normalization (Equation \ref{equ:norm}) before they are put into the CNN network. The permeability data needs to be logarithmic before normalization.
\begin{equation}
\label{equ:norm}
\displaystyle x^* = \frac{x - x_{min}}{x_{max} - x_{min}}
\end{equation}
where $x^*$ is the normalized value, $x$ is the value before normalization, $x_{min}$ and $x_{max}$ are the minimum and maximum values of all data before normalization, respectively. The normalized porosity and permeability values of all data are shown in Figure \ref{fig:KPorNormed}. It can be seen that the normalized porosity and normalized permeability have a linear relationship. The statistical distribution of normalized permeability and normalized porosity values are shown in Figure \ref{fig:distribution_afterNormed}. Like Figure \ref{fig:distribution_beforeNormed}(a), the samples of the normalized porosity values are evenly distributed. Compared with the distribution before normalization, the sample distribution of the normalized permeability values is more uniform. After the data is normalized, the training speed of the CNN model will be accelerated and it will be easier to converge \cite{Goodfellow-et-al-2016}. \par

\subsection{CNN architecture}

\begin{figure}[!htbp]
    \centering
	\includegraphics[width=0.8\linewidth]{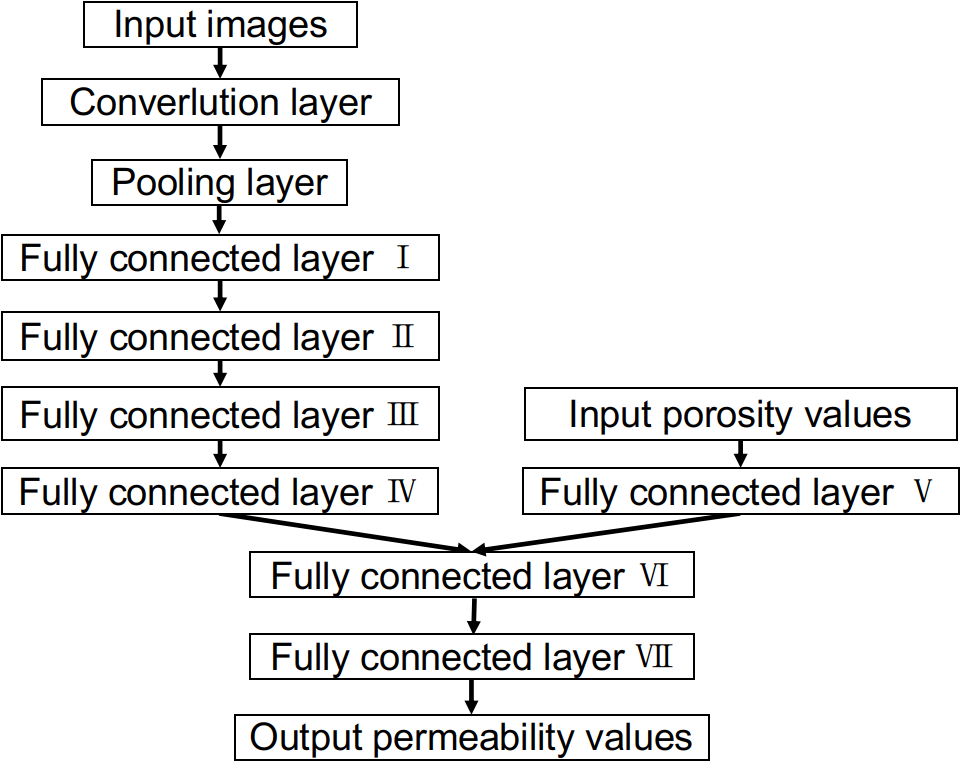}
	\caption{The architecture of CNN model.}
	\label{fig:MyCNN}
\end{figure}

After reference to the model in literatures \cite{Wu2019, Alqahtani2018} and adjustment of the model parameters, the architecture of our CNN model is shown in Figure \ref{fig:MyCNN}. Through the dot product of a filter and chunks of input data, the convolution layer generates a map of activations called a feature map, indicating the locations and strength of a detected feature in an input. The filter size of the convolution layer in this study is (2, 2). The Maximum Pooling method is used after the convolution layer. Its function is to progressively reduce the spatial size of the representation to reduce the number of parameters and computation in the network. Fully connected layers connect the input data processing layers with the output layer. The superposition of multiple fully connected lays will gradually reduce the high-dimensional data obtained from the pooling layer to the same dimension as the output data. The kernel size of the fully connected layers \RomanNumeralCaps{1} to \RomanNumeralCaps{7} are 128, 64, 16, 4, 1, 4 and 1, respectively. Activation functions are a critical part of the design of a neural network. The choice of activation function in the hidden layer will control how well the network model learns the training dataset. In this study, rectified linear activation (ReLU) is used as the activation function after the convolution layer and fully connected layers, which is the most common activation function used for hidden layers. The ReLU function is calculated as:
\begin{equation}
    ReLU(x) =\left\{\begin{matrix}
0.0, & for \quad x < 0\\ 
 x, & for  \quad x \geq 0
\end{matrix}\right.
\end{equation}

The dropout layer is used after the fully connected layer \RomanNumeralCaps{1} and \RomanNumeralCaps{2}. The Dropout layer is a mask that nullifies the contribution of some neurons towards the next layer and leaves unmodified all others. It offers a very computationally cheap and remarkably effective regularization method to reduce over-fitting and improve generalization error in deep neural networks of all kinds. The probability of retention in the dropout layer in this study is 0.5. Before the output layer, a neuron and linear activation function are used to reduce the data dimension of the fully connected layer to one dimension just as the output layer. \par
Mean squared error (MSE) is used as the loss function of this model, which is expressed as:
\begin{equation}
    loss = \frac{1}{n}\sum^n_{i=1}{y_i - y_i^*}^2
\end{equation}
where $y_i$ is the normalized permeability value, $y_i^*$ is the predicted value of normalized permeability. Adaptive moment estimation (Adam) is used as the optimizer of this model.\par
In this study, the pre-process of all data and the building of the CNN model are both based on the Python and TensorFlow platform.\par

\subsection{The training result}

\begin{figure}[!htbp]
    \centering
	\includegraphics[width=0.8\linewidth]{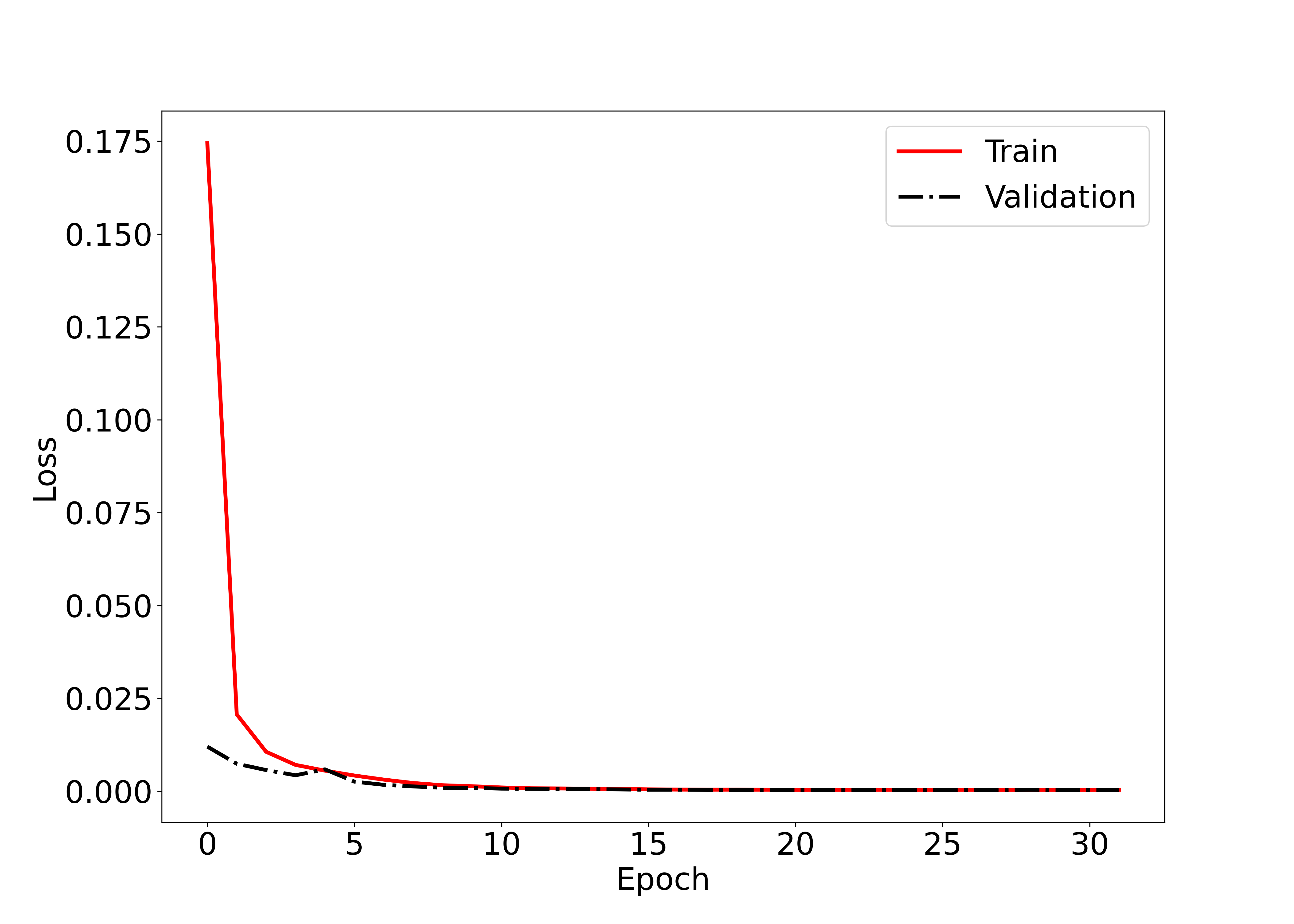}
	\caption{Loss values of training process.}
	\label{fig:loss}
\end{figure}

\begin{figure}[!htbp]
    \centering
	\includegraphics[width=0.8\linewidth]{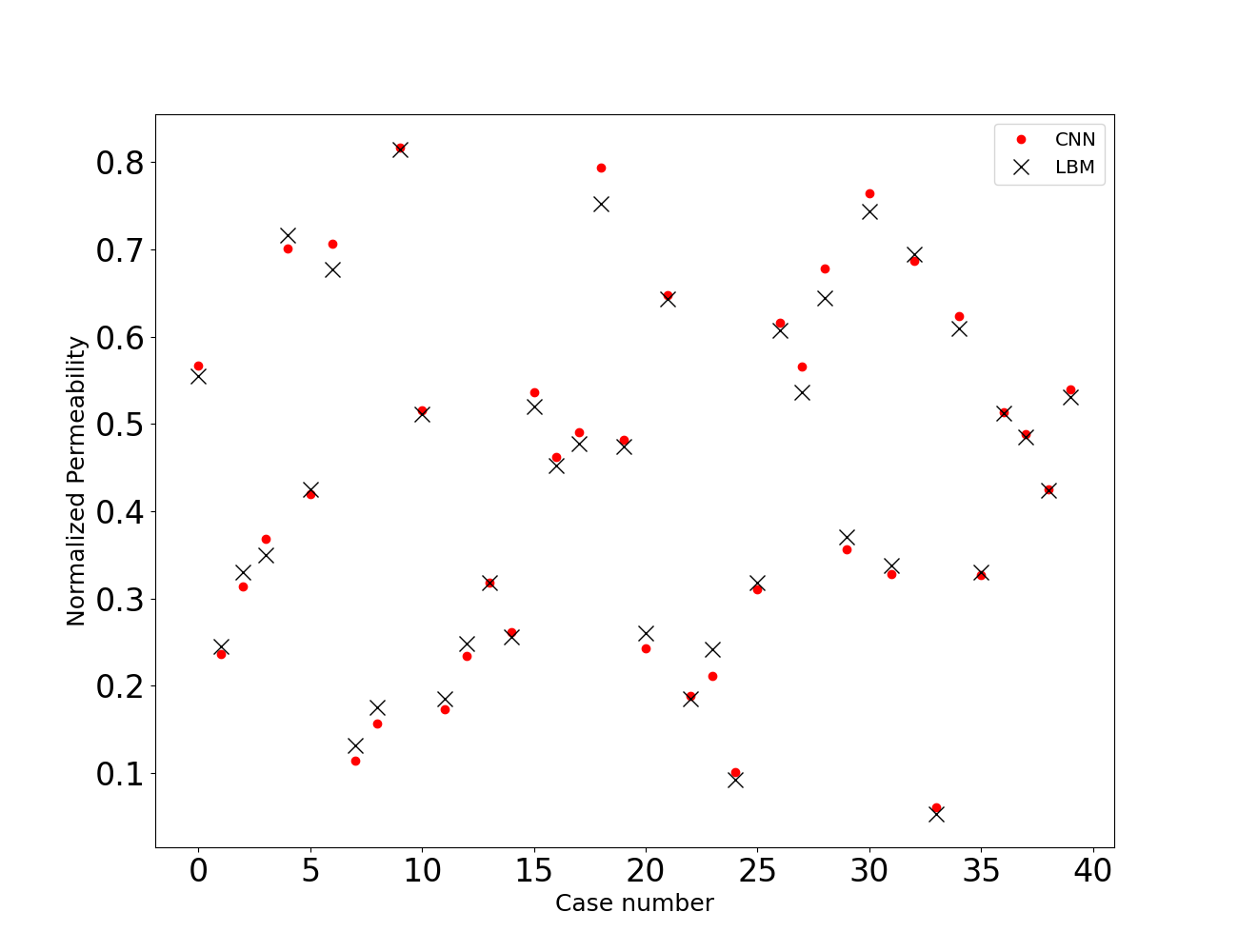}
	\caption{A comparison of 40 random samples in test dataset.}
	\label{fig:vs}
\end{figure}

Loss values of the training process are shown in Figure \ref{fig:loss}. It has been reduced to $3.06 \times 10^{-4}$ after 32 epochs of training. The comparison between the predicted and calculated permeability in the test dataset (300 samples) is carried out using the R square value (coefficient of determination),
\begin{equation}
\label{Equ:R2}
	R^2 = 1 - \frac{\sum_{i}(y_i - y^*_i)^2}{\sum_{i}(y_i - \bar{y})^2}
\end{equation}
where $\bar{y}$ is the average of all normalized permeability values in the test dataset. In the present study, the R square is 0.993. Compared with literature \cite{Alqahtani2020, Wu2019, Tian2020}, the prediction of our model is sufficiently accurate. A comparison of 40 random samples in the test dataset is presented in Figure \ref{fig:vs}, which also shows that the predicted and calculated values are in good agreement. \par
From the results obtained above, it can be concluded that the model we trained can be used for the permeability prediction of the porous media generated in section \ref{GoPM}. 

\section{Method}
\label{method}
In this section, the generation method of random packed isotropic porous media with specific porosity and permeability is detailed. \par

\begin{figure}[!htbp]
	\centering
	\begin{subfigure}{\textwidth}
		\centering
		\includegraphics[width=0.8\linewidth]{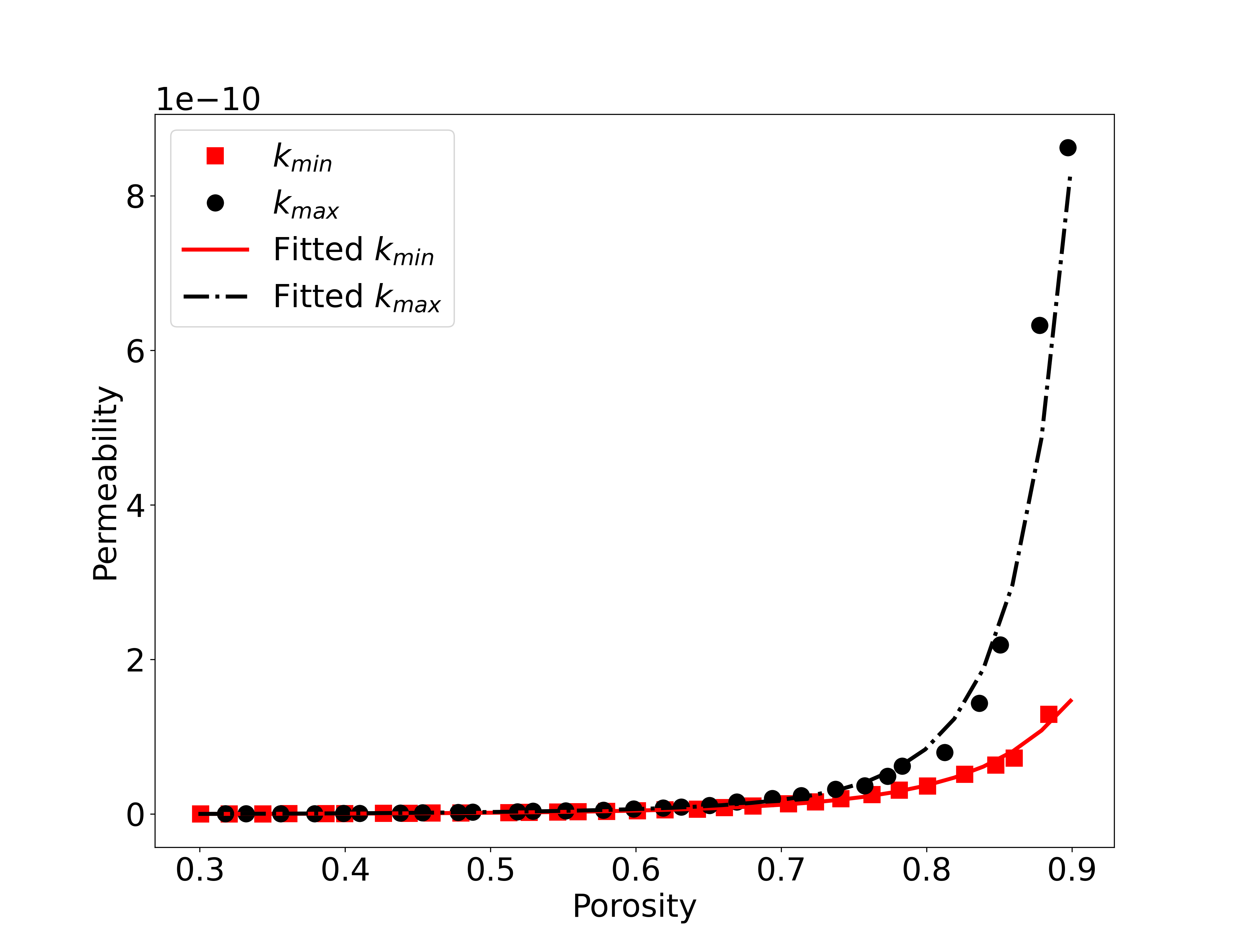}
		\caption{Raw values}
	\end{subfigure}
	\begin{subfigure}{\textwidth}
		\centering
		\includegraphics[width=0.8\linewidth]{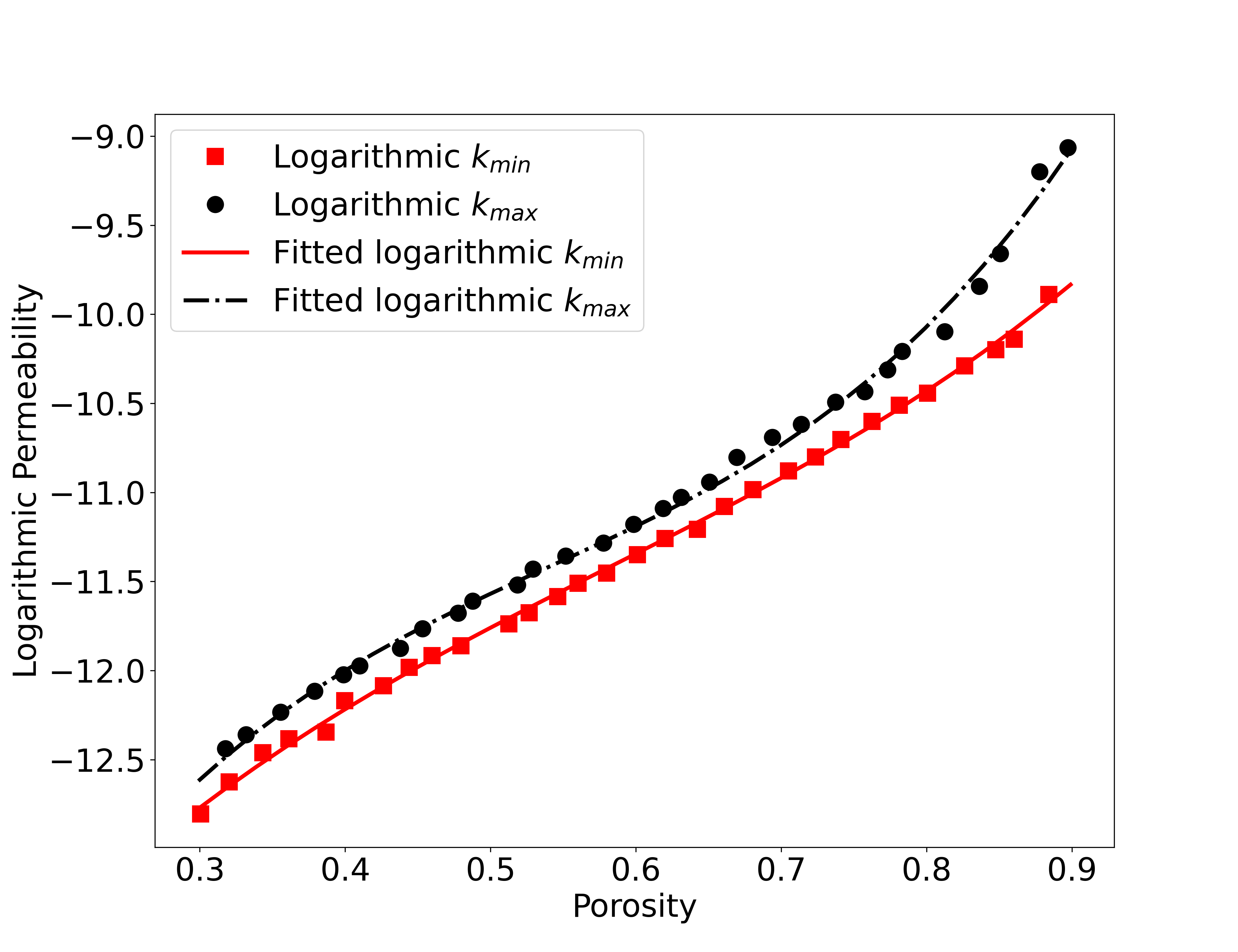}
		\caption{Normalized values}
	\end{subfigure}
	\caption{The relationship of the minimum and maximum permeability and porosity before and after logarithmization.}
	\label{fig:KPor2}
\end{figure}

Porous media with the same porosity may also have different permeability values, which can be seen from  Figure \ref{fig:KPor}.  Therefore, we take 0.02 as the interval of porosity, and pick the minimum and maximum logarithmic permeability values under each interval, and fit two curves in the form of equations as:
\begin{equation}
	\label{k_max}
	\log_{10}{k_{max}} = 21.63 \varphi^3 - 35.05 \varphi^2 + 22.66 \varphi - 16.84 \quad ,
\end{equation}
\begin{equation}
	\label{k_min}
	\log_{10}{k_{min}} = 8.744 \varphi^3 - 15.23 \varphi^2 + 12.95 \varphi - 15.52 \quad .
\end{equation}
Both the coefficients of determination of equations \ref{k_max} and \ref{k_min} are 0.99. Figure \ref{fig:KPor2} shows the comparison between the original data and the fitted data. From the two coefficients of determination and Figure \ref{fig:KPor2}, it can be concluded that the two fitting formulas can represent the upper and lower limits of the permeability value under a certain porosity.  \par

Based on the above trained model and fitted results of all data, the algorithm for generating a porous medium with a specific porosity and permeability is as follows. 
\begin{algorithm}
	\caption{ The generation of porous media with specific porosity and permeability}\label{algorithm2}
	\KwIn{porosity $\varphi$, the specified tolerance for porosity $\xi _{\varphi}$, permeability $k$, the specified tolerance for permeability $\xi _{k}$ }
	\KwResult{Porous media with porosity $\dot{\varphi}$ and permeability $\dot{k}$}
	Calculate the maximum and minimum permeability ($k_{max}$ and $k_{min}$) of porosity $\varphi$ with Equation \ref{k_max} and \ref{k_min}\;
	Make sure $k$ is between $k_{max}$ and ${k_{min}}$. If not, exit the program and re-enter the parameters\;
	\While{True}{
		With the method in Section \ref{GoPM}, generate a porous medium with porosity $\varphi$ and the specified tolerance for porosity $\xi_{\varphi}$, get the porosity $\dot{\varphi}$ of the generated one \;
		Calculate the permeability of generated porous media with trained CNN model, get the permeability $\dot{k}$\;
		\If{$|\dot{k} - k|/k < \xi_{k}$}{
			Exit loop\;
		}
	}
\end{algorithm}
\par
\begin{figure}[!htbp]
	\centering
	\includegraphics[width=0.8\linewidth]{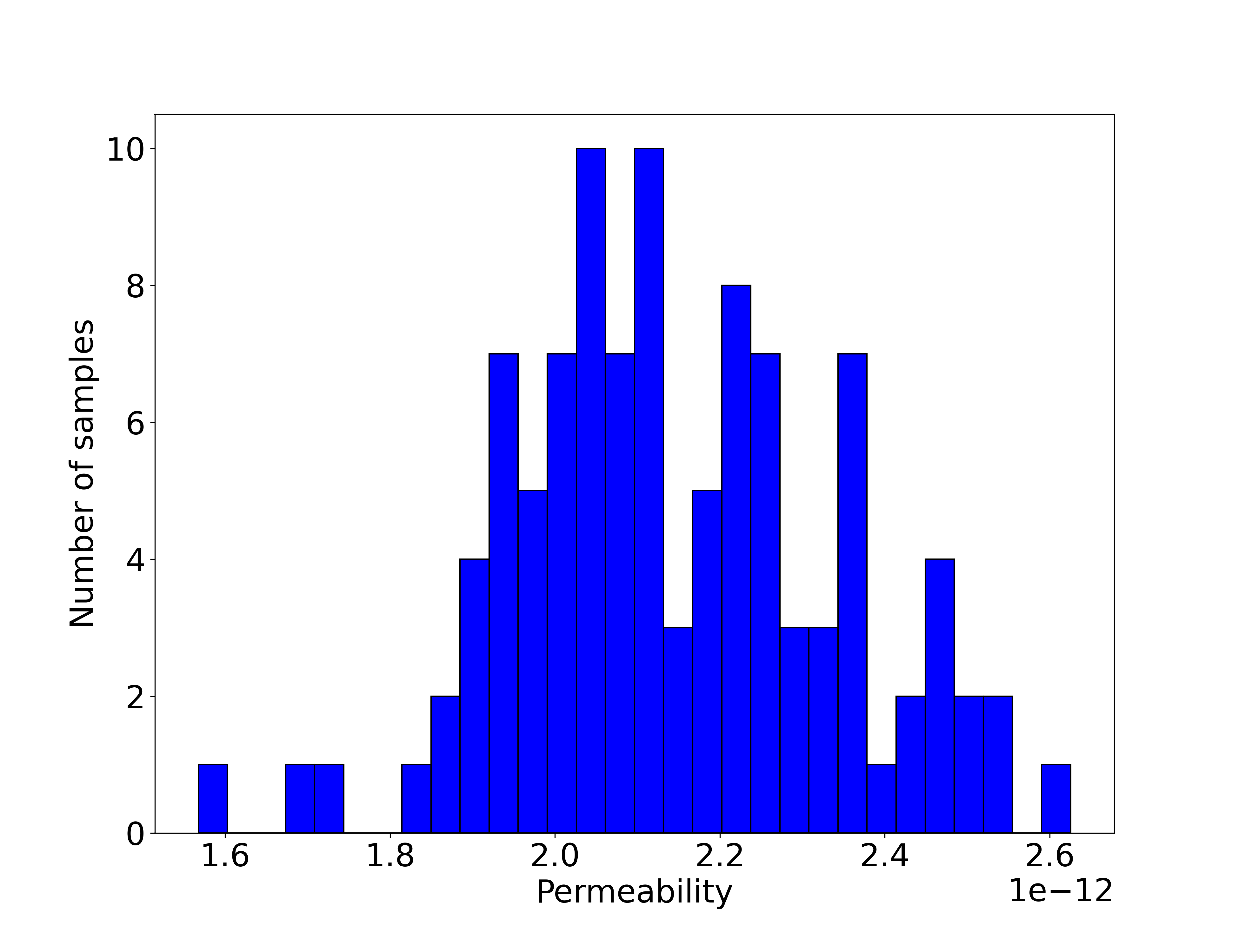}
	\caption{The permeability distribution of porous media generated around porosity $\varphi = 0.5$.}
	\label{fig:dis_0.5}
\end{figure}

This algorithm is a Monte Carlo method. In each cycle, the generated porous media and its permeability values are random. But if the porosity is specified, the permeability of the generated porous media must be between $k_{min}$ and $k_{max}$. In particular, we studied the distribution of the permeability of the porous media generated at a specific porosity. For example, the permeability distribution at $0.49 < \varphi < 0.51$ is shown in Figure \ref{fig:dis_0.5}. When $\varphi=0.5$, $k_{min} = 1.6\times10^{-12}$ and $k_{max} = 2.6\times10^{-12}$, which are consistent with the maximum and minimum values in Figure \ref{fig:dis_0.5}. It can be seen that the distribution of permeability approximately follows a normal distribution. The probability that the permeability of generated porous media lies near the value of $(k_{min}+k_{max})/2$ is the greatest for a specific porosity value. Therefore, selecting the target permeability according to the maximum probability will shorten the running time of this algorithm program.\par

For example, it takes about 1.8 seconds on average to generate a porous media on a single-threaded computer, on the condition of $\varphi = 0.55$, $\xi_{\varphi}=0.001$, $k=3.40\times 10^{-12}$,  $\xi_{k} = 0.01$. It will take less time to use multi-threading. \par

\section{Conclusion}
\label{conclusion}
In the present study, the porous media generation method with particle radius following lognormal distribution is proposed. The MRT-LBM-D2Q9 model is used to simulate the flow in porous media. Permeability can be calculated from the simulation. A CNN model, which can immediately predict permeability from a porous medium, is built and trained with 3000 samples. Compared with literature \cite{Alqahtani2020, Wu2019, Tian2020}, the prediction of our trained model is sufficiently accurate. Note that if the number of samples is larger, the statistics obtained would be more accurate and the training results of the model are expected to be even better. Based on the above implementations, a method for the generation of circle-packed isotropic porous media with specific porosity and permeability is proposed. Now, with the support of powerful computers, a porous medium that satisfies the error condition can be generated in a short time. \par
Future works will be devoted to the research about flow and heat transfer in a porous medium with specific porosity and permeability. This method will also be extended to three dimensions. 

\section{Acknowledgment}
\label{Acknowledgment}
The authors would like to thank the financial supports from NSFC (Grant No. 12172163, 12002148, and 11672124), the Shenzhen Peacock Plan  (KQTD-2016022620054656), and  Guangdong Provincial Key Laboratory of Turbulence Research and Applications (Grant No. 2019B21203001). This work is supported by Center for Computational Science and Engineering of Southern University of Science and Technology.



\bibliographystyle{model1-num-names}
\bibliography{ref}








\end{document}